\documentclass[epj]{svjour}
\usepackage{graphicx}
\graphicspath{{./Figures/}}
\usepackage{epstopdf}
\usepackage{textcomp}
\usepackage{lipsum}
\usepackage{dblfloatfix}
\usepackage{fixltx2e}
\usepackage{color}
\usepackage[square,comma,numbers,sort&compress]{natbib} 
\newcommand{\unige}{1}
\newcommand{\infnge}{2}
\newcommand{\hzdr}{3}
\newcommand{\tudresden}{4}
\newcommand{\infnpd}{5}
\newcommand{\unipd}{6}
\newcommand{\edinburgh}{9}
\newcommand{\infnna}{10}
\newcommand{\uniba}{7}
\newcommand{\lngs}{12}
\newcommand{\atomki}{13}
\newcommand{\infnto}{14}
\newcommand{\infnmi}{15}
\newcommand{\roma}{16}
\newcommand{\teramo}{19}
\newcommand{\gssi}{11}
\newcommand{\infnba}{8}

\newcommand{\budapest}{17}
\newcommand{\uniPi}{18}
\begin{document}
\title{A high-efficiency gas target setup for underground experiments, and redetermination of the branching ratio of the 189.5 keV $\mathbf{^{22}Ne(p,\gamma)^{23}Na}$ resonance}
\titlerunning{A high-efficiency gas target setup for underground experiments}
\author{
	F.\,Ferraro \inst{\unige,\infnge} \and
	M.P.\,Tak\'{a}cs\inst{\hzdr,\tudresden} \and
	D.\,Piatti \inst{\infnpd,\unipd} \and
	V.\,Mossa \inst{\uniba,\infnba} \and
	M.\,Aliotta \inst{\edinburgh} \and
	D.\,Bemmerer \inst{\hzdr}  \thanks{email: d.bemmerer@hzdr.de} \and
	A.\,Best \inst{\infnna} \and
	A.\,Boeltzig \inst{\gssi} \thanks{present address: University of Notre Dame and Joint Institute for Nuclear Astrophysics (JINA), Notre Dame, USA} \and
	C.\,Broggini \inst{\infnpd} \and
	C.G.\,Bruno \inst{\edinburgh} \and
	A.\,Caciolli \inst{\infnpd,\unipd} \thanks{email: caciolli@pd.infn.it} \and
	F.\,Cavanna\inst{\unige,\infnge} \and
	T.\,Chillery \inst{\edinburgh} \and
	G.\,F.\,Ciani \inst{\gssi,\lngs} \and
	P.\,Corvisiero \inst{\unige,\infnge} \and
	L.\,Csedreki \inst{\lngs} \and
	T.\,Davinson \inst{\edinburgh} \and
	R.\,Depalo \inst{\infnpd,\unipd} \and
	G.\,D'Erasmo \inst{\uniba,\infnba} \and
	A.\,Di Leva \inst{\infnna} \and
	Z.\,Elekes \inst{\atomki} \and
	E.\,M.\,Fiore \inst{\uniba,\infnba} \and
	A.\,Formicola \inst{\lngs} \and
	Zs.\,F\"ul\"op \inst{\atomki} \and
	G.\,Gervino \inst{\infnto} \and
	A.\,Guglielmetti \inst{\infnmi} \and
	C.\,Gustavino \inst{\roma} \and
	Gy.\,Gy\"urky \inst{\atomki} \and
	G.\,Imbriani \inst{\infnna} \and
	M.\,Junker \inst{\lngs} \and
	I.\,Kochanek \inst{\lngs} \and
	M.\,Lugaro \inst{\budapest} \and
	L.\,E.\,Marcucci \inst{\uniPi} \and
	P.\,Marigo \inst{\infnpd,\unipd} \and
	R.\,Menegazzo \inst{\infnpd} \and
	F.R.\,Pantaleo \inst{\uniba,\infnba} \and
	V.\,Paticchio \inst{\infnba} \and
	R.\,Perrino \inst{\infnba} \thanks{permanent address: INFN Sezione di Lecce, Lecce, Italy} \and
	P.\,Prati \inst{\unige,\infnge} \and
	L.\,Schiavulli \inst{\uniba,\infnba} \and
	K.\,St\"ockel \inst{\hzdr,\tudresden} \and
	O.\,Straniero \inst{\teramo} \and
	T.\,Sz\"ucs \inst{\atomki} \and
	D.\,Trezzi \inst{\infnmi} \and
	S.\,Zavatarelli \inst{\infnge}
	(LUNA\,collaboration)
	}                     
\authorrunning{F.\,Ferraro \textit{et al.} (LUNA collab.)}

\institute{
	Dipartimento di Fisica, Universit\`a degli Studi di Genova, Genova, Italy 
	\and 
	INFN Sezione di Genova, Genova, Italy 
	\and
	Helmholtz-Zentrum Dresden-Rossendorf (HZDR), Dresden, Germany 
	\and
	Technische Universit\"at Dresden, Dresden, Germany 
	\and
	INFN Sezione di Padova, Padova, Italy 
	\and
	Dipartimento di Fisica e Astronomia, Universit\`a di Padova, Padova, Italy 
	\and
	Dipartimento Interateneo di Fisica ``Michelangelo Merlin", Universit\`a degli Studi di Bari, Bari, Italy 
	\and
	INFN, Sezione di Bari, Bari, Italy 
	\and
	SUPA, School of Physics and Astronomy, University of Edinburgh, Edinburgh, United Kingdom 
	\and
	Dipartimento di Fisica "E. Pancini", Universit\`a di Napoli Federico II and INFN, Sezione di Napoli, Strada Comunale Cinthia, 80126 Napoli, Italy 
	\and
	Gran Sasso Science Institute, L'Aquila, Italy 
	\and
	INFN, Laboratori Nazionali del Gran Sasso, Assergi, Italy 
	\and 
	Institute of Nuclear Research (MTA Atomki), Debrecen, Hungary 
	\and
	Dipartimento di Fisica, Universit\`a di Torino, and INFN Sezione di Torino, Torino, Italy 
	\and 
	Universit\`a degli Studi di Milano and INFN Sezione di Milano, Milano, Italy 
	\and
	INFN Sezione di Roma ``La Sapienza", Roma, Italy, 
	\and
	Konkoly Observatory, Research Centre for Astronomy and Earth Sciences, Hungarian Academy of Sciences, 1121 Budapest, Hungary 
	\and
	Dipartimento di Fisica ``E. Fermi", Universit\`a di Pisa, and INFN Sezione di Pisa, Pisa, Italy 
	\and
	INAF Osservatorio Astronomico di Teramo, Via Mentore Maggini, 64100 Teramo, Italy 
}
\date{Draft \today }
%
\abstract{
The experimental study of  nuclear reactions of astrophysical interest is greatly facilitated by a low-background, high-luminosity setup. The Laboratory for Underground  Nuclear Astrophysics (LUNA) 400\,kV accelerator offers ultra-low cosmic-ray induced background due to its location deep underground in the Gran Sasso National Laboratory (INFN-LNGS), Italy, and high intensity, 250-500\,$\mu$A, proton and $\alpha$ ion beams.
In order to fully exploit these features, a high-purity, recirculating gas target system for isotopically enriched gases is coupled to a high-efficiency, six-fold optically segmented bismuth germanate (BGO) $\gamma$-ray detector. 
The beam intensity is measured with a beam calorimeter with constant temperature gradient.
Pressure and temperature measurements have been carried out at several positions along the beam path, and the resultant gas density profile has been determined.
Calibrated $\gamma$-intensity standards and the well-known $E_p$ = 278\,keV $\mathrm{^{14}N(p,\gamma)^{15}O}$ resonance were used to determine the $\gamma$-ray detection efficiency and to validate the simulation of the target and detector setup.
As an example, the recently measured resonance at $E_p$ = 189.5 keV in the $^{22}$Ne(p,$\gamma$)$^{23}$Na reaction has been investigated with high statistics, and the $\gamma$-decay branching ratios of the resonance have been determined.
\PACS{
	{26.30.-k}{Nucleosynthesis in novae, supernovae, and other explosive stars} \and
	{25.40.Ep}{Inelastic proton scattering} \and
	{29.30.Kv}{X- and gamma-ray spectroscopy} \and
	{25.40.Lw }{Radiative capture}\and
	{25.40.Ny }{Resonance reactions}\and
	{26.20.Cd }{Stellar hydrogen burning}
}
} 
\maketitle

\section{Introduction}
\label{sec:Intro}
The $\mathrm{^{22}Ne(p,\gamma)^{23}Na}$ reaction, which belongs to the NeNa cycle, is active in high temperature hydrogen burning \cite{Iliadis15-Book}. One astrophysical site of particular interest for this reaction is the  Hot Bottom Burning (HBB) process in asymptotic giant branch (AGB) stars of high initial mass (M $>$ 4 M$_\odot$), which are one of the proposed candidates to explain Na anomalies in ancient stellar globular clusters  \cite{Gratton04-ARAA}. 

The rate of this reaction is controlled by a large number of resonances. Despite recent experimental work on resonances in the 400-1200 keV range \cite{Longland10-PRC,Depalo15-PRC}, the rate was still highly uncertain especially due to the contribution of  low-lying  resonances. As a result, the recommended median rates from two widely used thermonuclear reaction rate compilations, NACRE \cite{NACRE99-NPA} and Iliadis \cite{Iliadis10-NPA841_31}, differed by two to three orders of magnitude, especially at the temperatures relevant for the HBB process. 

In order to provide a more accurate estimate, the reaction was recently studied at the Laboratory for Underground Nuclear Astrophysics (LUNA) 400\,kV accelerator, 
using a windowless gas target experiment and two large high-purity germanium detectors. In this campaign, three predicted resonances located at 156.2, 189.5, and 259.7 keV in the laboratory system were observed for the first time, and their energies and strengths determined \cite{Cavanna15-PRL,Depalo16-PRC}. 
Strength values for two of the three newly found resonances  were found to be much larger than previous indirect upper limits \cite{Iliadis10-NPA841_31}, confirming the need for direct nuclear-reaction measurements.

The full implications of the resultant, revised thermonuclear reaction rate have yet to be explored. Initial work exploring thermally pulsing AGB stars experiencing the HBB process suggests a larger amount of $^{23}$Na ejected \cite{Slemer17-MNRAS}.
Very recently, the observation of two of the three new resonances has been independently confirmed \cite{Kelly17-PRC}, albeit with slightly larger strengths and somewhat different decay branching ratios. 
The present work puts the necessary parts in place to push the study of the $\mathrm{^{22}Ne(p,\gamma)^{23}Na}$ reaction to ultra-low energies. To this end, a new setup was devised to achieve a hundredfold higher efficiency than the previous one at LUNA \cite{Cavanna14-EPJA,Cavanna15-PRL,Depalo16-PRC}. The purpose of this new setup is to search for two  proposed resonances at very low energy, $E_p$ = 71 and 105 keV, not observed yet 
\cite{Cavanna15-PRL,Depalo16-PRC}, and the direct capture contribution.

Section \ref{sec:Setup} introduces the new high-efficiency setup, including its complete characterisation. The background observed in the $\gamma$-ray detector, both with and without incident ion beam, is analysed in Section \ref{sec:Background}. As a demonstration of the capabilities of the new setup, the decay branching ratios of the $E_p$ = 189.5\,keV resonance in  \\ $\mathrm{^{22}Ne(p,\gamma)^{23}Na}$ are determined  in Section \ref{sec:189.5keVres}. 
Finally, a summary and outlook are given in Section \ref{sec:Summary}.


\section{Experimental setup}
\label{sec:Setup}
The LUNA 400 kV electrostatic accelerator is located deep underground in the INFN Gran Sasso National Laboratory (LNGS), Italy. It provides $^1$H$^+$ or $^4$He$^+$ ion beams with high currents, up to 250-500 $\mathrm{\mu A}$ on target, with very small momentum spread and excellent long-term stability \cite{Formicola03-NIMA}. 

Experimental results obtained at LUNA have been reviewed elsewhere \cite{Costantini09-RPP,Broggini10-ARNPS,Broggini18-PPNP}.

\begin{figure*}[tbh]
\centering
\includegraphics[angle=0,width=\textwidth]{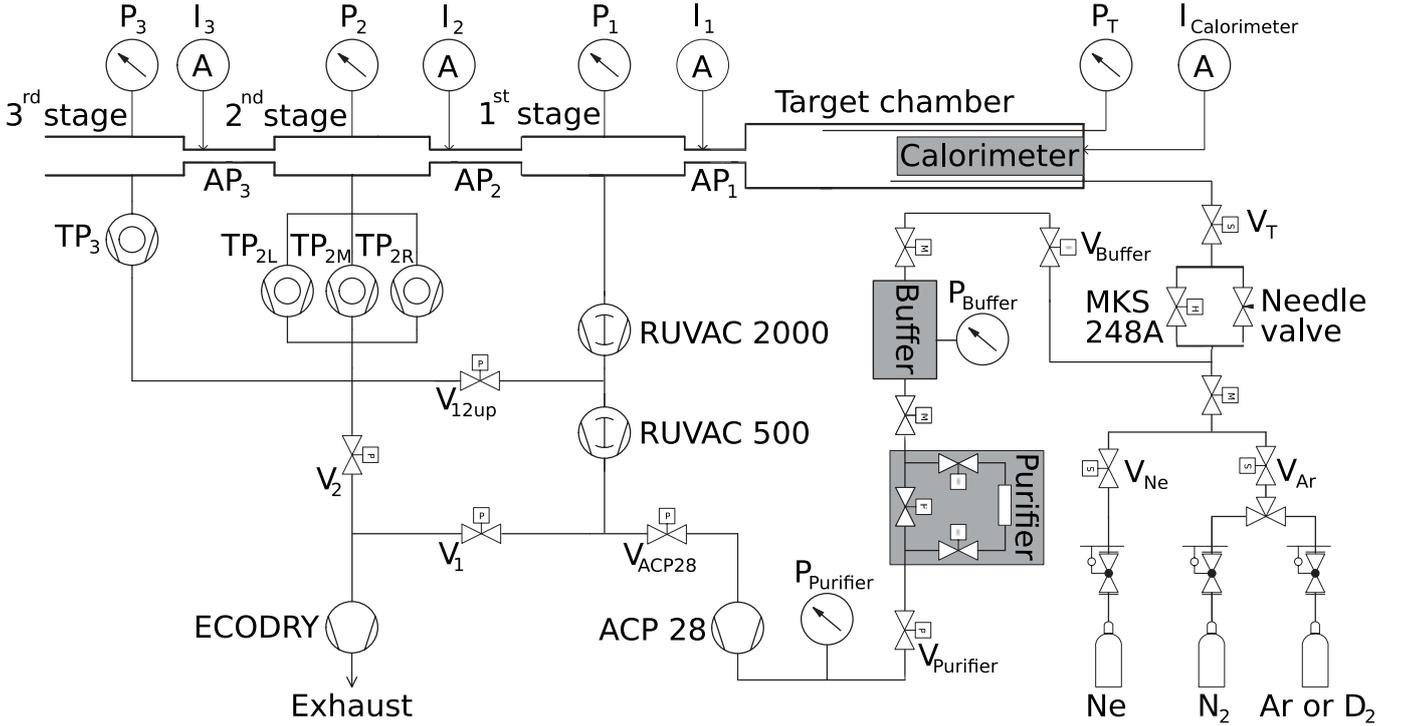}
\caption{Schematic diagram (not to scale) of the differential pumping system. In recirculating mode, the valves \texttt{V1}, and \texttt{V2} are closed. The beam comes from the accelerator on the left, passes through the apertures \texttt{AP$_{\mbox{3}}$}, \texttt{AP$_{\mbox{2}}$} and \texttt{AP$_{\mbox{1}}$}, enters the target chamber and is stopped in the calorimeter. More than 99.5\% of the gas, which enters the chamber close to the calorimeter, is pumped through the RUVAC 2000. Approximately 0.5\% of the gas still flows towards the second pumping stage and a negligible part flows in the third pumping stage.}\label{fig:GasTarget}
\end{figure*}

\subsection{General considerations}

The experimental setup is optimised for the irradiation of target materials that exist in gaseous form at normal temperature and pressure. The design considerations have been guided by three nuclear reactions in particular: \\$^{22}$Ne(p,$\gamma$)$^{23}$Na, $^{22}$Ne($\alpha$,$\gamma$)$^{26}$Mg, and $^{2}$H(p,$\gamma$)$^{3}$He. These studies entail the use of isotopically enriched $^{22}$Ne (9\% abundance in natural neon) and $^2$H (0.012\% abundance in natural hydrogen).

Due to the hindering effect of the repulsive Coulomb barrier at the astrophysically relevant beam energies accessible at LUNA, the nuclear reactions under study exhibit ultra-low cross sections, which  drop exponentially  with decreasing beam energy. At the same time, these energies lie near the  maximum in the stopping power curve \cite{Ziegler10-NIMB}, the so-called Bragg peak.  

This combination of low, and rapidly varying, cross section and high stopping power requires a careful optimisation of the experimental conditions. As a result, chemical compounds \cite[e.g.]{Caciolli12-EPJA} or implanted targets \cite[e.g.]{Depalo15-PRC} are disfavored, because they contain a different nuclear species in addition to the nucleus under study, further enhancing the stopping power but not the experimental yield. Gas cells \cite[e.g.]{Bordeanu13-NPA} are problematic, as well, because their entrance windows may cause unwanted beam energy straggling or  even stop a low-energy ion beam altogether. 

\subsection{Windowless gas target}

The solution adopted here is a windowless, extended gas target of the static type (Figure \ref{fig:GasTarget}). The  gas enters the target chamber from the right (\texttt{VT} in Figure \ref{fig:GasTarget}), with the incoming gas flux precisely regulated by the \texttt{MKS248A} valve controlled by a pressure measurement device, to keep the target pressure constant within 0.5\%. 

The fact that there is no entrance window leads to some inevitable gas loss through the metal tube functioning as target entrance collimator (\texttt{AP$_{\mbox{1}}$} in Figure \ref{fig:GasTarget}). This effect is mitigated by limiting the diameter of \texttt{AP$_{\mbox{1}}$} to 7\,mm, and by making it relatively long, 40\,mm. 

The target gas is then removed from the setup by  large Roots-type vacuum pumps  (\texttt{RUVAC 2000} and \texttt{RUVAC 500} in Figure \ref{fig:GasTarget}), which have a pumping speed of 2050 and 505 m$^3$/h, respectively, over a relatively wide pressure range. The \texttt{RUVAC 2000}, its vacuum recipient, and a connecting tube matching it to \texttt{AP$_{\mbox{1}}$} form the first pumping stage. For typical target pressures of 2.0 (0.3) mbar in the $^{22}$Ne ($^2$H$_2$) case, the pressure in the first pumping stage is found to be almost two orders of magnitude lower with respect to the pressure inside the target.

The combination of a long, narrow collimator and a powerful pump is  repeated twice, for the second (collimator \texttt{AP$_{\mbox{2}}$} and turbomolecular pumps \texttt{TP$_{\mbox{2L}}$, TP$_{\mbox{2M}}$, TP$_{\mbox{2R}}$}, two of them with 1000 l/s and one with 1500 l/s nominal pumping speed, respectively) and third pumping stages (collimator \texttt{AP$_{\mbox{3}}$} and turbomolecular pump \texttt{TP$_{\mbox{3}}$} with 350 l/s nominal speed). After the third pumping stage, the connecting conditions to the accelerator (pressure in the 10$^{-6}$ mbar range and negligible gas flow) are met.

When employing $^{22}$Ne gas, it is necessary to re-use the gas exhaust from the three pumping stages in order to limit consumption. To this end, the exhausted gas is collected, compressed, and guided to a chemical getter (Monotorr PS4-MT3-R-2 with a PS4-C3-R-2 heated getter) to remove impurities, typically oxygen, nitrogen, hydrogen, water, oxocarbons, and hydrocarbons.  The cleaned gas is then transported to a buffer (volume 1 liter, typical pressure 400-700 mbar) and then re-used as input gas.

In addition to $^{22}$Ne gas, other gases are needed  for calibration and background study purposes. In order to study the ion beam induced background, natural argon gas is used. 
For the determination of the detection efficiency at high $\gamma$-ray energies, it is possible to insert nitrogen gas to exploit the $E_p$ = 278\,keV $^{14}$N(p,$\gamma$)$^{15}$O resonance. 

The core of the setup, the gas target chamber, is a stainless steel tube designed to fit inside a $\gamma$-ray calorimeter formed by a 4$\pi$ bismuth germanate (BGO) detector  (see Section \ref{subsec:BGO}). A new chamber has been designed for the present experiment and has been characterised as described in the following sections. In addition to the  chamber, a  calorimeter, which is different with respect to the one used in the previous experiment \cite{Cavanna14-EPJA,Cavanna15-PRL,Depalo16-PRC}, was installed to monitor the beam current. Its characterisation is described in section \ref{subsec:Calorimeter}. 

In order to monitor the system performance, pressure sensors are connected to the target chamber by a long copper tube, to each of the three pumping stages, and to the buffer and purifier. The status of pumps and valves is controlled by a LabVIEW system and logged together with the pressure values. Typical pressures observed during the experiment, with 2 mbar of neon in the target chamber, were: from 8$\cdot 10^{-2}$ to 4$\cdot10^{-3}$~mbar in the first stage, 1.5$\cdot10^{-6}$~mbar in the second stage, and 1$\cdot10^{-7}$~mbar in the third stage. Similar ratios were observed also for different values of the pressure in the scattering chamber. The first stage is the closest to the scattering chamber, therefore a particular attention was devoted to monitor the pressure  inside that part as discussed in section \ref{sec:pressure}.
The most critical aperture is of course  \texttt{AP$_{\mbox{1}}$}, where a compromise between the required drop in pressure and the beam spatial dimension has to be taken into account. Selecting an aperture of 7 mm, we obtained more than one order of magnitude reduction in pressure (see section \ref{sec:pressure} for details) and less then 5\% of the total beam current deposited on the collimator, typically. The other two apertures are less restrictive and the current deposited on them is lower than the 1\% of the total beam current.

At the high beam intensities of LUNA, direct water cooling is necessary for all the three collimators (\texttt{AP$_{\mbox{1}}$} to \texttt{AP$_{\mbox{3}}$}), so these are kept at a temperature of 13\,$^\circ$C (286\,K, measured with a PT100 thermistor connected to \texttt{AP$_{\mbox{1}}$}). The outer walls of the target chamber and the pumping stages are at room temperature,  22\,$^\circ$C.

\subsection{Pressure, temperature, and density profile}
\label{sec:pressure}
\begin{figure*}[tb]
\includegraphics[width=\columnwidth]{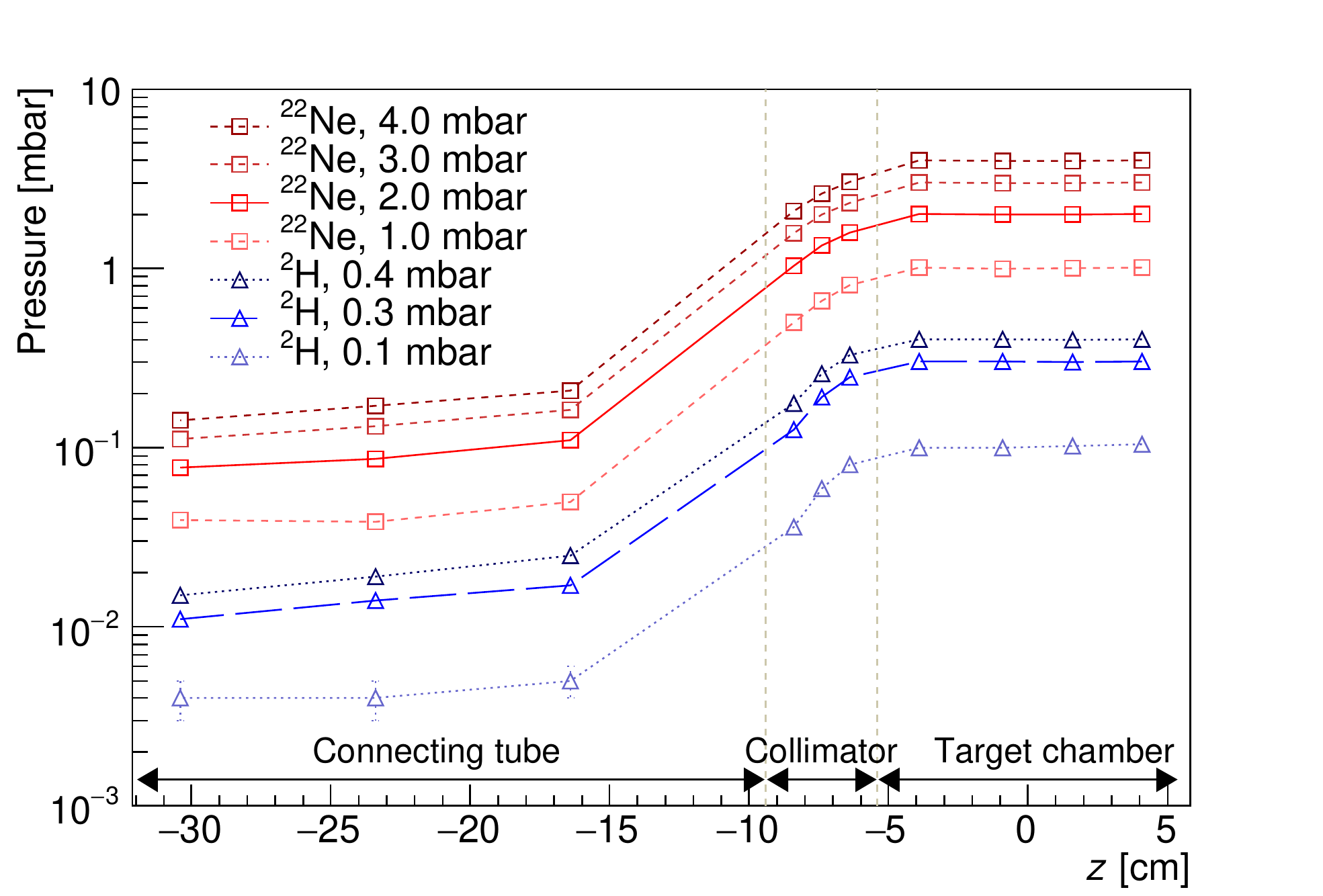}%
\includegraphics[width=\columnwidth]{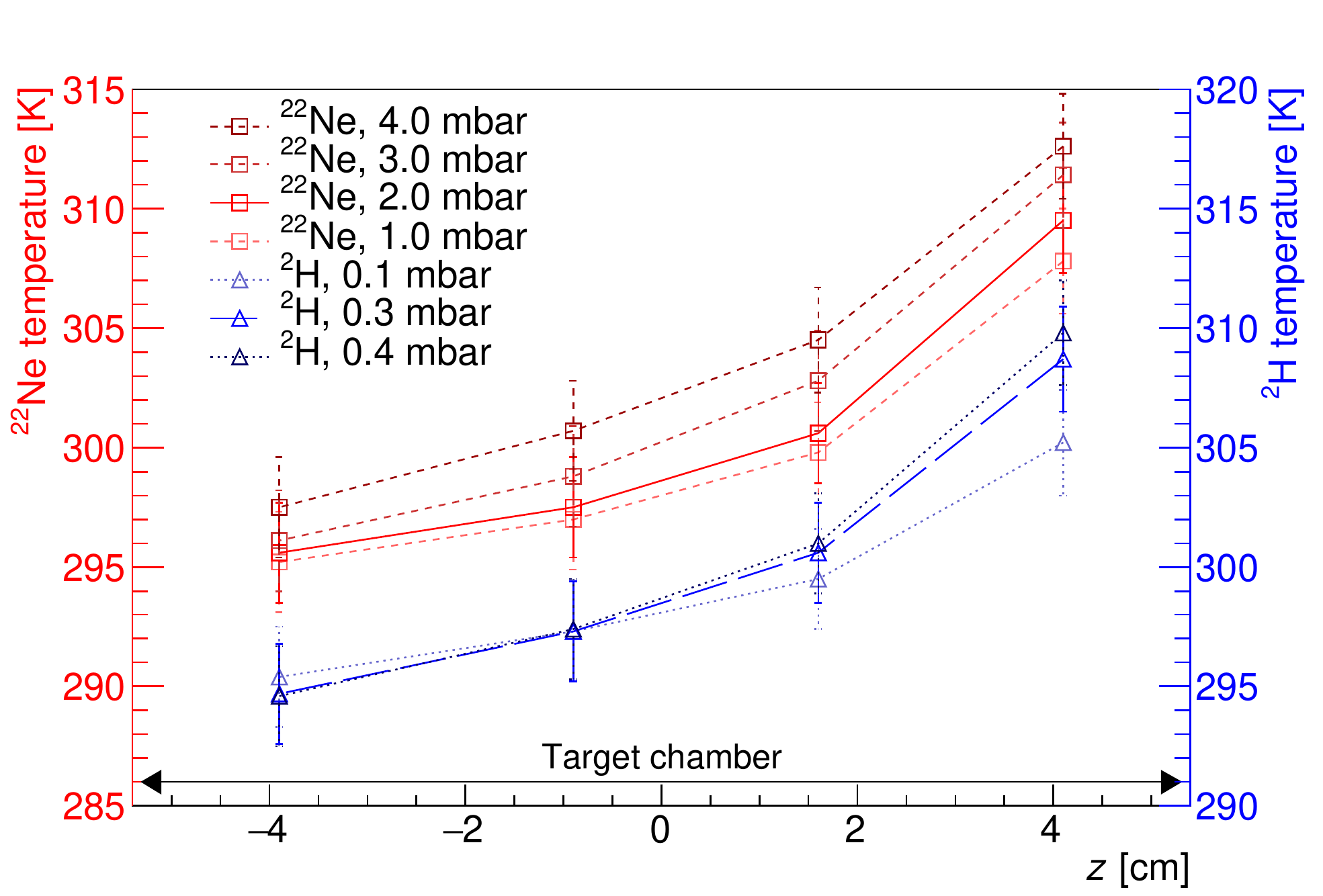}
\caption{
\label{fig:TemperatureProfile}
\label{fig:PressureProfile}
Left panel: Measured pressure profile $p(z)$. The center of the chamber is placed at $z = 0$ cm, while the calorimeter surface corresponds to $z = 5$ cm.   --- 
Right panel: Temperature profile. In order to improve the readability, the temperature profile for $\mathrm{^2H}$ gas is displayed with respect to the right axis, which is shifted by 5 K. The position $z$ is measured from the center of the target chamber. On both panels, the lines, connecting the points, are not the results of any fit and they have been tracked just to guide the reader's eye.}
\end{figure*}

A precise understanding of the gas target density profile (i.e. the gas density $n(z)$ as a function of position $z$ along the beam axis) is needed for two reasons. First, for  resonance yield measurements included in the $^{22}$Ne(p,$\gamma$)$^{23}$Na and $^{22}$Ne($\alpha$,$\gamma$)$^{26}$Mg studies, the gas density determines the beam energy loss and thus the position inside the target chamber where the maximum yield of the resonance is reached. This position, in turn, is needed because of the position dependence of the $\gamma$-ray detection efficiency. Second, for the analysis of the non-resonant cross section included in the $^{22}$Ne(p,$\gamma$)$^{23}$Na and $^{2}$H(p,$\gamma$)$^{3}$He studies, the density must be known in order to properly normalise the yield.

For the determination of the density profile $n(z)$, the temperature $T(z)$ and pressure $p(z)$ were measured at a number of positions $z$ inside the target chamber and in the connecting tube between collimator \texttt{AP$_{\mbox{1}}$} and the main recipient of the first pumping stage. These measurements were performed with precise copies of the target chamber and connecting tubes, which had respectively KF10 and KF16 vacuum ports directly connected  to them to enable the use of pressure and temperature sensors.

For the pressure profile, four calibrated capacitance-type pressure gauges (two MKS Baratron 626 and two Pfeiffer CMR 363, typical precision 0.3\%) were used to measure the pressure at  ten different positions (shown from right to left in Figure \ref{fig:PressureProfile}, left panel): four inside the target chamber, three in the \texttt{AP$_{\mbox{1}}$} collimator and three positions in the connecting tube. The collimator measurements were enabled by thin tubes (internal diameter 0.5\,mm) fixed at the sides of a specially prepared copy of \texttt{AP$_{\mbox{1}}$}. The gauges were changed in position between measurement ports to connect the data points and to check the consistency of the pressure calibrations of the various gauges. 

The pressure measurements were then repeated for eight different target pressures in the 0.5-4.0 mbar range for $^{22}$Ne and for seven  in the 0.1-1.0 mbar range for $^2$H. The overall behaviour is the same for all nominal target pressures studied; selected profiles are shown in Figure \ref{fig:PressureProfile}, left panel.  

It is found that the pressure inside the target chamber is constant to $\pm$0.5\%. Inside the 40\,mm long, 7\,mm narrow collimator \texttt{AP$_{\mbox{1}}$}, there is a monotonic decrease of the target pressure, as expected for a high-impedance tube. Inside the connecting tube, the trend continues but with lower slope, consistent with the fact that the 100\,mm wide connecting tube is significantly larger than \texttt{AP$_{\mbox{1}}$}, and thus the tube has a significantly lower gas flow impedance. The final uncertainty for the pressure inside the target chamber is $\pm$0.9\%, taking into account calibration, reproducibility, and profile.

The gas temperature was only measured inside the target chamber  (Figure \ref{fig:PressureProfile}, right panel), using four PT100 thermistors. 
In the small  watercooled collimator \texttt{AP$_{\mbox{1}}$} the gas is in close contact with the watercooled surfaces and  we have  therefore assumed in the computation a temperature of 286 K.
However, considering the gas amount inside the collimator (18\% of the total), an error of 10\,K on its temperature would cause a 0.6\% error in the gas density.
Inside the connecting tube, it is assumed that the temperature of the gas is the same as  the outside temperature of the tube (295\,K).

Inside the target chamber, the temperature drops monotonically between the beam stop (heated to 343\,K, see Section \ref{subsec:Calorimeter} below) and \texttt{AP$_{\mbox{1}}$} (cooled to 286\,K). A measurement very close to the beam stop was not performed due to the large solid angle covered by the beam stop and, hence, significant radiative heating of the PT100 sensors. The temperature profile inside the chamber is determined with 0.5\% relative uncertainty (1.5 K). 

For the temperatures in the collimator and connecting tube, 1\% uncertainty (3 K) is conservatively assumed. For any given incident ion, the effective target thickness observed by the beam in the connecting tube and collimator is always less than 30\% of the total gas thickness, so that this effect contributes an additional uncertainty of 0.3\% for the total gas thickness.

Using the pressure and temperature data $p(z)$ and $T(z)$, the gas density $n(z)$ was then calculated using the ideal gas law%
\begin{equation}\label{eqn:IdealGasLaw}
n(z) = \frac{\nu N}{V} = \frac{\nu \cdot p(z)}{k_B \cdot T(z)}
\end{equation}
where $N$ is the number of gas molecules per volume $V$, $\nu$ the number of  atoms per molecule of gas ($\nu$ = 1 for neon, $\nu$ = 2 for deuterium), and $k_B$  the Boltzmann's constant.

\begin{figure}[tbh]
\includegraphics[width=\columnwidth]{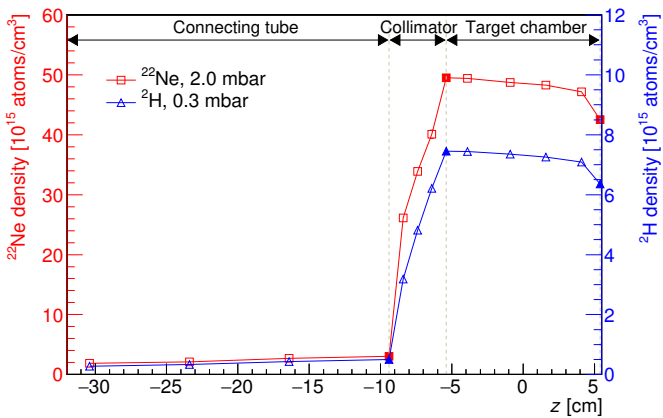}
\caption{\label{fig:DensityProfile}
Calculated density profile from Equation (\ref{eqn:IdealGasLaw}). The density close to the calorimeter surface and the points on the dashed vertical lines (filled squares and triangles) have been extrapolated as described in the text. The profile for $\mathrm{^2}$H gas uses the right axis, which has five times lower range than the left axis. The position $z$ is measured from the center of the target chamber.}
\end{figure}

For the parts where no data are available, i.e., close to the edge of each of the three segments (connecting tube, collimator, and target chamber) the density is extrapolated linearly based on  measured pressure and density profiles inside the segment. 
For the interface between collimator and connecting tube, the two extrapolations do not match perfectly. 
The value from the connecting tube extrapolation is adopted but the discrepancy is included in full in the error budget, entailing a 0.7\% uncertainty in the integrated gas thickness. Figure \ref{fig:DensityProfile} shows the resultant density profile for the adopted working pressures of 2.0\,mbar for the $^{22}$Ne(p,$\gamma$)$^{23}$Na campaign  and of 0.3\,mbar for the $^{2}$H(p,$\gamma$)$^{3}$He one.

Taking into account the uncertainties from the pressure and temperature measurements and the extrapolation, a total uncertainty of 1.3\% is found for the integrated gas thickness. The same error is also adopted to each individual density measurement.

Finally, an intense ion beam may lead to some thinning of the gas target, by the so-called beam heating effect \cite{Goerres80-NIM,Marta06-NIMA}. The beam-heating correction in neon gas was studied previously using the resonance scan technique \cite{Cavanna14-EPJA}, but in a much larger chamber. Using those measurements, for a beam intensity of 250\,$\mu$A at a beam energy of $E_p$ = 100\,keV, a correction of 8\% is found for 2\,mbar $^{22}$Ne gas and of 0.9\% for 0.3\,mbar $^2$H gas. However, the present chamber is narrower, so the conductive cooling of the heated gas volume is more efficient.
Taking this effect into account \cite{Osborne84-NPA}, a correction of 6\% (0.6\%) is found for the 2\,mbar $^{22}$Ne gas (0.3\,mbar $^2$H gas) case. For the correction, a relative uncertainty of 20\% is adopted, which is included for each run based on the actual beam intensity and added in quadrature to the above mentioned 1.3\%.

\subsection{Beam calorimeter}
\label{subsec:Calorimeter}

When the target chamber is filled with gas, a beam intensity measurement with a Faraday cup becomes impractical due to secondary electrons, therefore a different approach is followed here, using a power compensation calorimeter \cite{Vlieks83-NIM,Casella02-NIMA}. 

\begin{figure}[tb]
\includegraphics[width=\columnwidth]{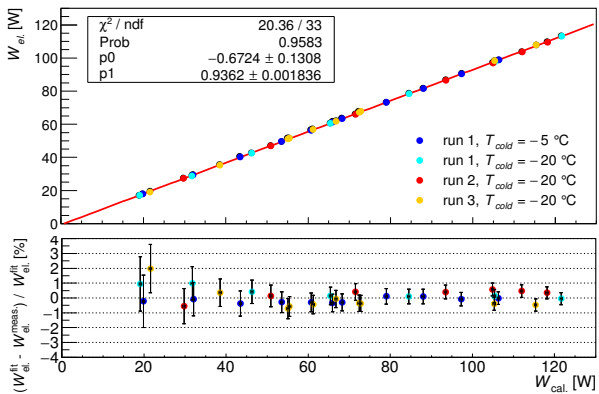}
\caption{\label{fig:CaloCalib}
Top panel: electrical calibration of the calorimeter to determine the parameters of Eq.~(\ref{eq:Wel_Wcalo}). The statistical error bars are smaller than the size of the data points. -- Bottom panel: residuals. See text for details.}
\end{figure}

The calorimeter is made from copper and consists of three parts: the hot side (70 $^\circ$C, acting also as the beam stop), the heating resistors, and the cold side (at different possible temperatures as reported in Figure \ref{fig:CaloCalib}). The hot and cold sides are always kept at constant temperatures by regulating the current through the heating resistors for the hot side, and  the cooling power in a feedback-controlled chiller for the cold side. Thus, there is always a constant temperature gradient between the hot and the cold sides.

The beam stop can be heated up either by the resistors or by the ion beam, thus, the more power is provided by the beam, the less is provided by the resistors.
If $W_{0}$ is the power delivered by resistors while the beam is off and $W_{\rm run}$ is the power delivered when the beam is on, the calorimetric beam power is given by 
\begin{equation}
W_{\rm cal} = W_0 - W_{\rm run}.
\end{equation}

The calorimetric power values $W_0$ and $W_{\rm run}$ are calculated by measuring the voltage and current for the heating resistors.
Two dividers were designed and used to decouple the power circuit from the readout, similarly to what reported in a previous work \cite{Casella02-NIMA}: 
The voltage divider is made of a passive resistive series with high resistance (3 $\times$ 33 k$\mathrm{\Omega}$ resistors), so that any possible influence of the voltmeter on the power circuit is negligible.
The current divider is a LEM LAH 25-NP current transducer, completely decoupled from the power circuit.
The outputs from the two dividers are then measured by a NI-cRIO-9207 module and logged by the LabVIEW control software.
The latter, together with NI-cRIO controller and modules, actively controls the calorimeter operations and logs the data every second \cite{Ferraro17-PhD}.

The statistical uncertainty on the calorimetric reading of the power was found to be 0.4 W, based on the ripple and stability of the calorimeter readings. 

Before the beam intensity can be obtained, the $W_{\rm cal}$ value must be electrically calibrated by associating it to the electrical power $W_{\rm el}$ (measured using the chamber and calorimeter as a Faraday cup) using the equation
\begin{equation}
W_{\rm el} = p_0 + p_1 W_{\rm cal} \label{eq:Wel_Wcalo}.
\end{equation}
The two parameters  $p_0$ and $p_1$ reflect the facts that parasitic currents may lead to a slight overestimate of the calorimetric heat power, and that the heat flow from the hot side is very similar, but not completely equal, for the cases of localised heating by the beam and of more spread out heating by the resistors.

In order to experimentally determine $p_0$ and $p_1$, a dedicated setup, without gas in target chamber, was used.
A copper ring was mounted inside the target chamber, electrically insulated both from the chamber and the collimator and biased to -300 V, in order to suppress secondary electrons generated on the calorimeter surface. The electrical current impinging on the calorimeter was then integrated over 180 s with a calibrated current integrator and averaged. 
The beam, passing through the residual gas in the target chamber ($<$10$^{-3}$mbar), ionises the gas and some positive charges are collected by the above mentioned ring. This small positive current ($\sim$1-3\% of the current in the Faraday cup) measured on the ring is therefore added to the current. The final electrical current is then compared to the average $W_{\rm cal}$ value over the same time period (Figure \ref{fig:CaloCalib}).
Four calibration data sets were taken: one with a chiller setting of $-5$\,$^\circ$C and three more   at $-20$\,$^\circ$C. The results were found to be consistent (Figure \ref{fig:CaloCalib}) and averaged. The ion beam current is  finally given by
\begin{eqnarray}\label{eq:BeamCurrent}
I & = & \frac{p_0 + p_1 W_{\rm cal}}{\left(E_p - \Delta E_{p}^{\rm cal}\right)} \times e, \\
{\rm with}  \nonumber \\ 
p_0 & = & (-0.67 \pm 0.13) \; {\rm W}, \nonumber \\
p_1 & = & (0.936 \pm 0.002), \;  \nonumber
\end{eqnarray}
where $E_p$ is the beam energy and $\Delta E_{p}^{\rm cal}$ is the energy loss of the beam when passing through the target gas (i.e., the full target thickness including connecting tube, collimator, and target chamber), and $e$ the electric charge. 

Taking into account the calorimeter uncertainties, the error on the electrical reading, and the calibration, the final uncertainty on the beam intensity $W_{\rm el}$ is 1.5\% or 0.5\,W, whichever is larger.

\begin{figure*}[tb]
\centering
\includegraphics[width=0.25\textwidth]{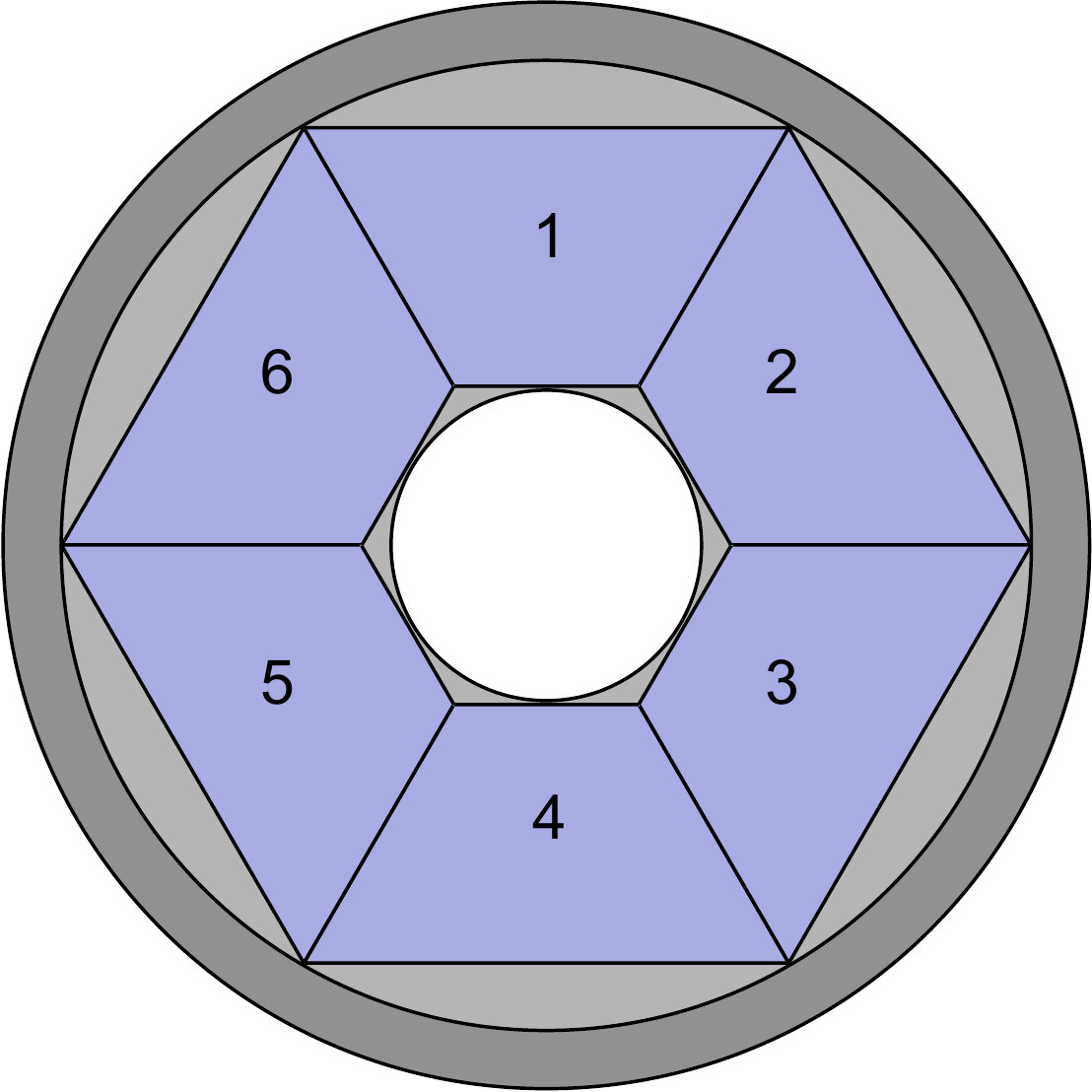}
\includegraphics[width=0.66\textwidth]{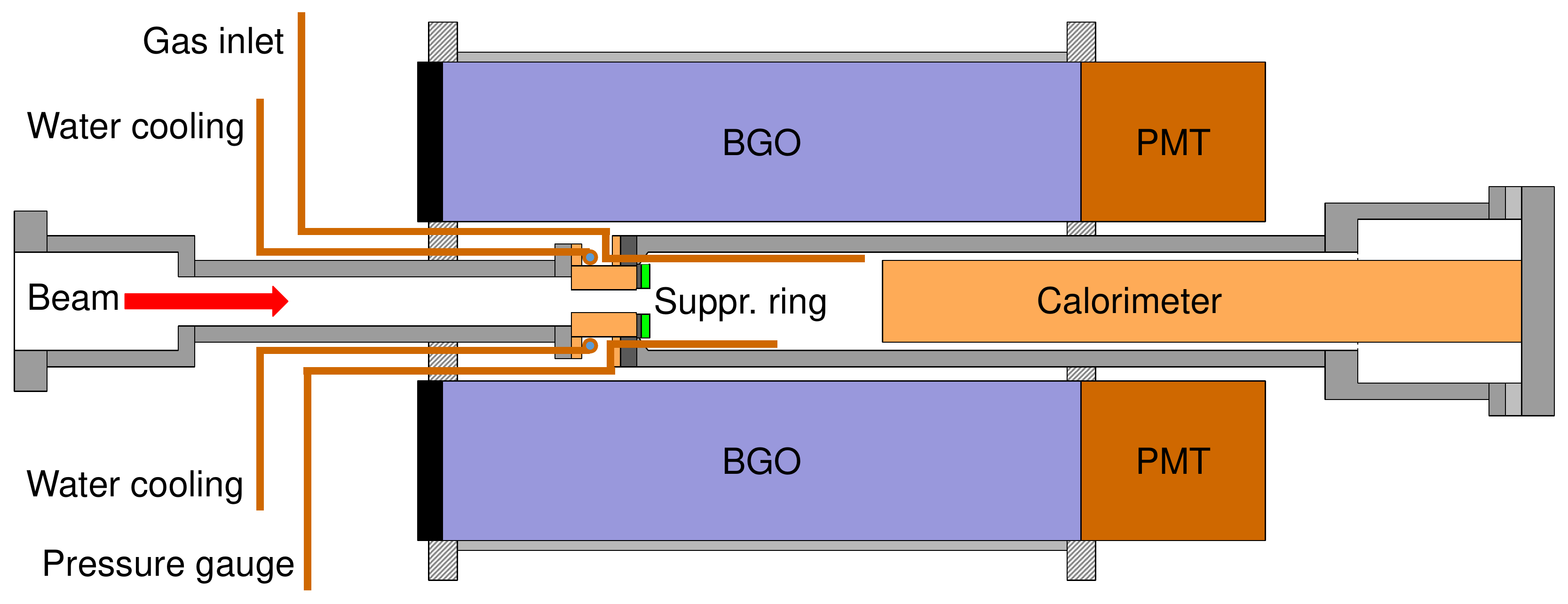}
\caption{\label{fig:BGO_front}
Left panel: Front cross section of the BGO detector. --- Right panel: Lateral cross-section view of the BGO. Also the target chamber and calorimeter are shown. The ring used only during the calorimeter calibrations and made from copper is also shown.}
\end{figure*}

\subsection{BGO and DAQ}
\label{subsec:BGO}

For the detection of the emitted $\gamma$ rays, an optically segmented bismuth germanate (BGO) borehole detector was used, the same as in previous work \cite{Caciolli11-AA}. 

The detector is composed of six scintillating crystals, each 28 cm long and 7 cm thick at the thinnest point, arranged in a hexagonal configuration surrounding the interaction volume (the resolution of each segment is about 11\% at 1.33 MeV \cite{Boeltzig17-JOP}). They are housed inside a stainless steel casing fitted with a borehole of 6\,cm diameter (Figure \ref{fig:BGO_front}).
Each crystal is covered with a reflecting foil, except for the opening for the photomultiplier tube (PMT, Hamamatsu R1847-07).
A CAEN V6533P high voltage power supply supplies an individual high voltage to each PMT, and the voltage was adjusted to match the gains of the individual PMTs.

\begin{figure}[tbh]
\includegraphics[width=\columnwidth]{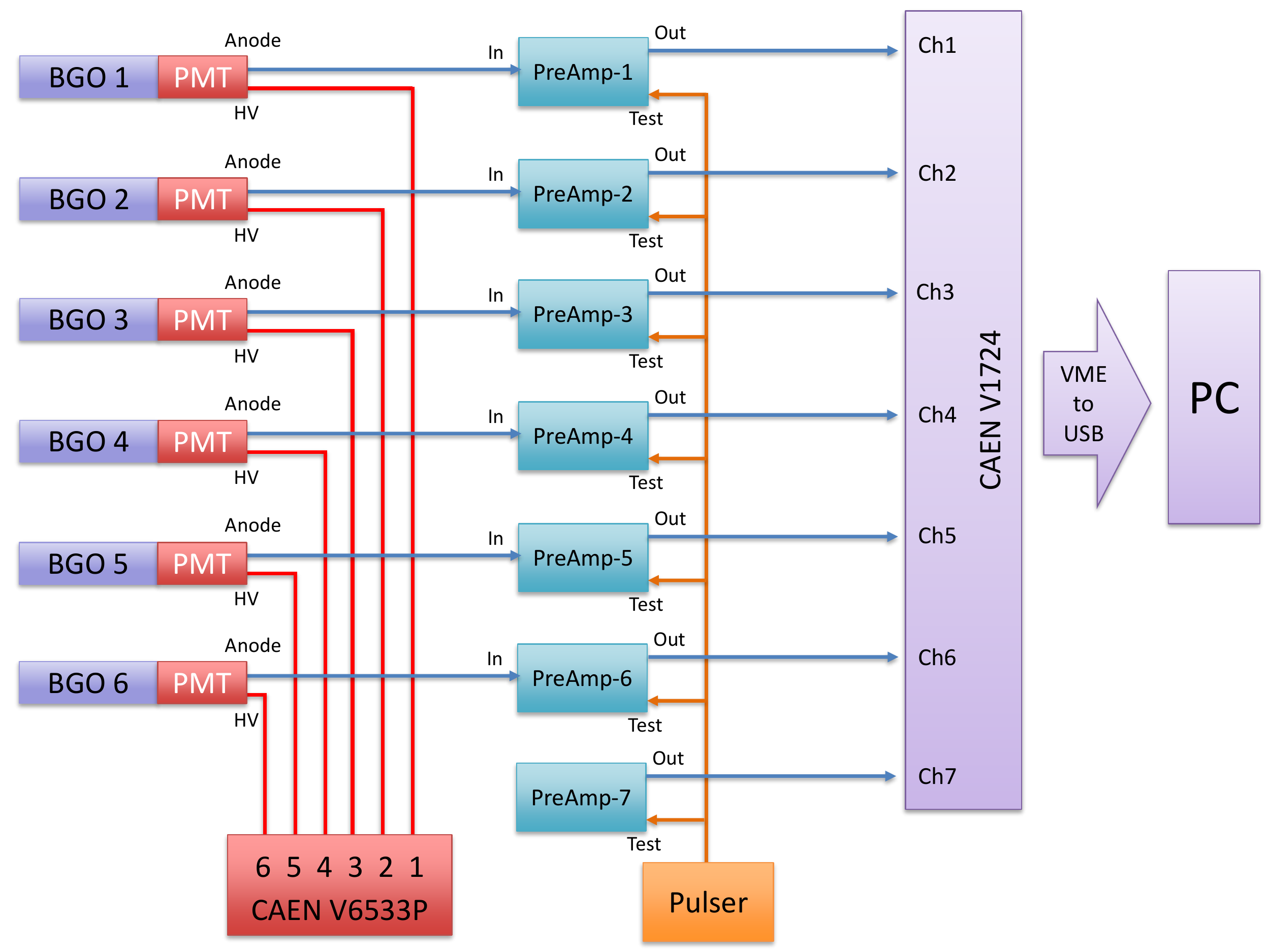}
\caption{\label{fig:Electronics}
Electronics scheme. See text for details.}
\end{figure}

For each of the six PMTs, the anode signal is passed to an Ortec 113 preamplifier. A pulse generator with approximately 50\,Hz rate, 100 ns pulse length, is connected to each of the test inputs of the six preamplifiers, and to a seventh preamplifier used to monitor the performance of the pulser. The preamplifier output of each segment is then connected to a CAEN V1724 (8 channel, 14 bit, 100 MS/s) digitiser, hence to the PC by a USB interface (see Figure \ref{fig:Electronics}). Each of the digitiser channels triggers independently, and the charge is integrated in the preamplifier. A trapezoidal filter is then applied to determine the height of the preamplifier signal and this information is stored, together with its time stamp, for offline analysis \cite{Boeltzig17-JOP}. 

The gain of the individual channels is determined run by run from the three most prominent laboratory background peaks: the 1.461 MeV $\gamma$ ray from $^{40}$K decay, the 2.204 MeV $\gamma$ ray from $\mathrm{^{214}Bi}$, and the 2.615 MeV $\gamma$ ray from $^{208}$Tl. The data are then sorted to events using a conservatively chosen coincidence time window of 3.5\,$\mu$s and stored as ROOT trees \cite{ROOT}.

In the offline analysis, first the dead time is determined for each individual channel by comparing the number of events in the pulser peak of that channel with the number of events in the seventh, pulser-only channel, which is assumed to be dead time free. Second, the pulser events are removed from the BGO channels by gating out all events where an event is recorded in the pulser-only channel (channel 7). 

Two general types of spectra are then created: First, a so-called add-back spectrum using the energies from all segments summed together, as if the BGO were one single detector. Second, a so-called singles sum spectrum formed by simply summing the individual histograms \cite{Boeltzig17-JOP}. 

The linearity of the gain calibrations was verified using the high-energy $\gamma$ rays from the $\mathrm{^{14}N(p,\gamma)^{15}O}$ reaction.

\subsection{$\gamma$-ray detection efficiency}
\label{subsec:Efficiency}

The $\gamma$-ray detection efficiency is measured using $\mathrm{^7Be}$, $\mathrm{^{137}Cs}$, $\mathrm{^{60}Co}$, and $\mathrm{^{88}Y}$  point-like radioactive sources, calibrated to better than 1\%, at 9 positions along the beam axis inside the interaction chamber. To achieve this, a special source holder was designed made of light materials to limit self-absorption.

\begin{figure}[tbh]
\includegraphics[width=\columnwidth]{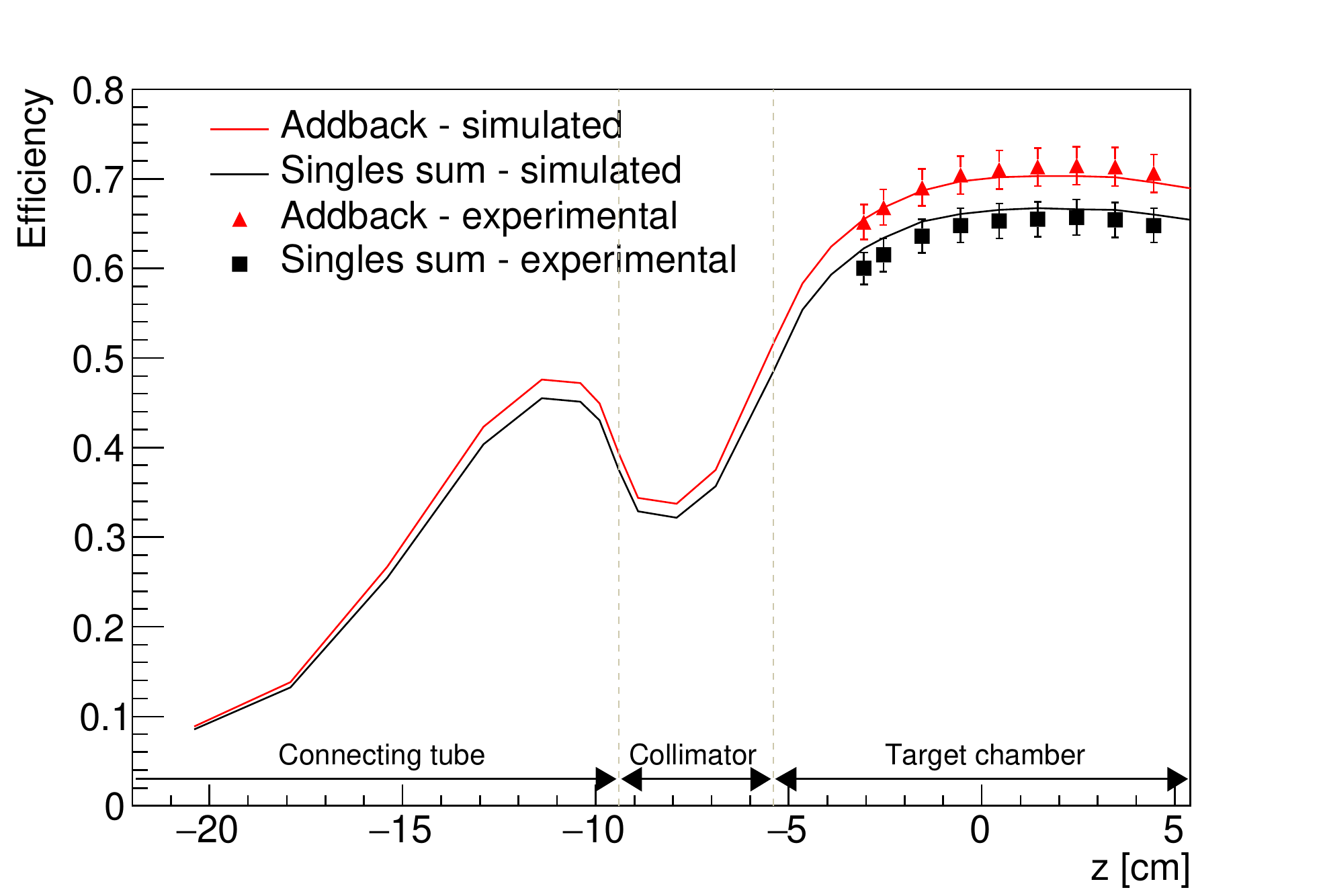}
\caption{\label{fig:137Cs_Simulation/Experiment}
Efficiency profile along the detector axis for a calibrated $^{137}$Cs source, compared with the Geant4 simulation. The position $z$ is measured from the center of the target chamber.}
\end{figure}

In addition, a GEANT4 Monte Carlo simulation was developed to determine the detection efficiency at positions and energies that are inaccessible with the sources. The simulation was found to match the experimental efficiency for the radioactive sources within 4\% without any rescaling (Figure \ref{fig:137Cs_Simulation/Experiment}). The simulation reproduces also the additional passive layers due to the cooling system of the collimator as shown by the reduction in efficiency from $z$ = -5 cm to $z$ = -10 cm in Figure  \ref{fig:137Cs_Simulation/Experiment}. In this region it was not possible to measure the efficiency experimentally, but, as shown clearly in Figure \ref{fig:DensityProfile}, the density drops by one order of magnitude in this region bringing to a negligible value its contribution to the experimental yield.
As a consistency check, the setup was also described using the well-tested LUNA Geant3 code \cite{Casella02-NIMA}, with consistent results.

\begin{figure}[tbh]
\includegraphics[width=\columnwidth]{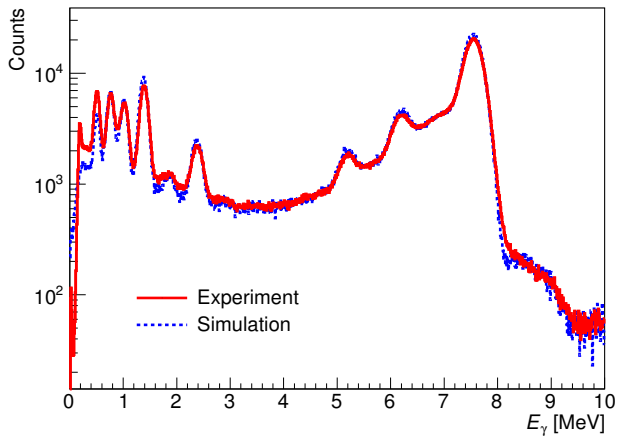}
\caption{\label{fig:14N_Simulation/Experiment}
$\mathrm{^{14}N(p,\gamma)^{15}O}$ add-back spectrum taken at the resonance energy $E_p$ = 278 keV. Comparison between the experimental spectrum and the spectrum simulated using the GEANT4 code.}
\end{figure}

For the ratio of high-energy to low-energy response, the simulations were validated using the peaks originating from the well known $\mathrm{^{14}N(p,\gamma)^{15}O}$ reaction, and a good agreement was found (Figure \ref{fig:14N_Simulation/Experiment}). 

For the present setup, the uncertainty associated to the validation of the simulations has been assumed to be 4\%.

\section{Background}
\label{sec:Background}

\begin{figure}[tbh]
\includegraphics[width=\columnwidth]{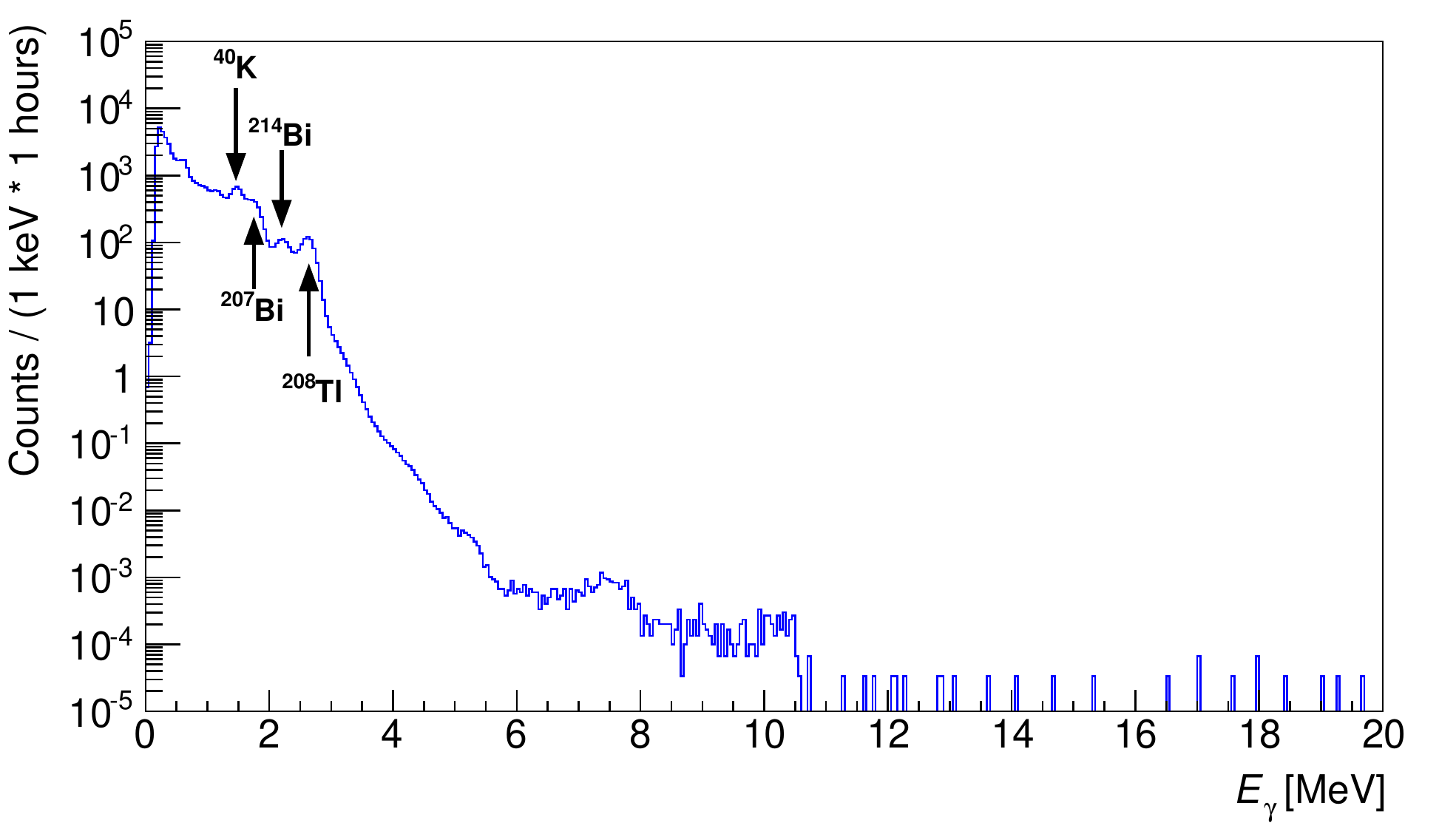}
\caption{\label{fig:NaturalBackground}
Experimental natural background spectrum. }
\end{figure}

Below 3 MeV $\gamma$-ray energy, natural background dominates the observed add-back spectrum (Figure \ref{fig:NaturalBackground}). This background is actually used as a tool to determine the energy calibration for each individual run (see Section \ref{subsec:BGO} above).

In laboratories above the Earth's surface, the cosmic-ray induced background above 3 MeV usually plays a critical role at low counting rate. However, LUNA benefits from a 1400 m thick rock overburden ($\approx$ 3800 meter water equivalent), which reduces the muon flux by a factor of $10^6$ and the neutron flux by a factor of $10^3$ \cite{Costantini09-RPP,Broggini10-ARNPS}. Still, a natural background contribution in the region from 5.5 to 10.5 MeV remains due to  (n,$\gamma$) reactions with the detector and experimental setup as discussed in details in \cite{Boeltzig17-JOP,Bemmerer05-EPJA}.

In add-back mode, the BGO detector, thanks to its high efficiency and the large solid angle coverage, is able to effectively detect the $\gamma$ rays originating from the decay of the excited state of the final nucleus, resulting in a peak at $Q+E$ (where $Q$ is the reaction $Q$-value and $E$ is the center of mass energy at which the reaction takes place)  in the add-back spectrum. For the reactions under consideration here, $Q$ is always larger than 3\,MeV, namely $Q$ = 8.794 MeV for $^{22}$Ne(p,$\gamma$)$^{23}$Na, 5.493 MeV for $^{2}$H(p,$\gamma$)$^{3}$He, and 10.615 MeV for $^{22}$Ne($\alpha$,$\gamma$)$^{26}$Mg, so that in all cases $Q+E>$ 3\,MeV, and full advantage is taken of the cosmic-ray suppression at LUNA, as shown by the ultra-low rate observed in these energy regions without beam (Figure \ref{fig:NaturalBackground}).

Because of this unique situation, another source of background needs careful attention, namely background produced by the ion beam. Indeed, nuclear reactions involving light contaminants may produce high energy $\gamma$ rays, leading to a background which depends on the contaminants in the setup, their position, and the beam energy \cite{Bemmerer05-EPJA}.
In case of proton beam, the most relevant reactions for ion-beam induced background found in the present setup are $\mathrm{^{11}B(p,\gamma)^{12}C}$, $\mathrm{^{14}N(p,\gamma)^{15}O}$, $\mathrm{^{15}N(p,\gamma)^{16}O}$, $\mathrm{^{19}F(p,\alpha\gamma)^{16}O}$ and $\mathrm{^{18}O(p,\gamma)^{19}F}$.

The $\mathrm{^{19}F(p,\alpha\gamma)^{16}O}$ reaction dominates the spectrum for beam energies above its $E_p$ = 340\,keV resonance, with a very large  peak at 6.13\,MeV due to the decay of the second excited state of $^{16}$O. The $\mathrm{^{18}O(p,\gamma)^{19}F}$ reaction is problematic near its $E_p$ = 151\,keV resonance, because its $Q$-value of 7.994 MeV is only 1\,MeV lower than the $^{22}$Ne(p,$\gamma$)$^{23}$Na reaction, and it may in some cases require a narrowing of the $\gamma$-ray region of interest in the data analysis. 

Above $E_p$ = 278 and 300 keV the \linebreak $\mathrm{^{14}N(p,\gamma)^{15}O}$ (\cite{Lemut06-PLB,Bemmerer06-NPA}, $Q$ = 7.297\,MeV) and $\mathrm{^{15}N(p,\gamma)^{16}O}$ (\cite{Bemmerer09-JPG,Caciolli11-AA}, $Q$ = 12.127\,MeV) reactions contribute to the background, leading to sum peaks at 7.6 MeV and 12.4 MeV, respectively.

Finally and most importantly, the $\mathrm{^{11}B(p,\gamma)^{12}C}$ was found to contribute significantly to the counting rate, not only at its $E_p$ = 163\,keV resonance but also above and even below, favoured by the relatively low atomic number of boron. The signature of this reaction are $\gamma$-rays at 16.1, 11.7, and 4.4 MeV, and  a significant additional background from Compton scattering in the region of interest for the studied reactions.

In order to subtract the ion-beam induced background from the $^{22}$Ne(p,$\gamma$)$^{23}$Na and $^{22}$Ne($\alpha$,$\gamma$)$^{26}$Mg energy spectra, monitor runs with an inert noble gas, natural argon, were performed. For these runs, the argon gas pressure was set so that the energy loss with argon was the same as with neon gas, in order to mimic also the lateral size of the beam and thus hit similar parts of the target chamber. Similarly, helium gas was used to monitor the beam induced background for the $^2$H(p,$\gamma$)$^3$He reaction.


\section{Decay branching ratios of the $E_p$ = 189.5 keV resonance in $^{22}$Ne(p,$\gamma$)$^{23}$Na}
\label{sec:189.5keVres}

With this setup, a new study of the recently discovered \cite{Cavanna15-PRL,Depalo16-PRC} and confirmed \cite{Kelly17-PRC} $\mathrm{^{22}Ne(p,\gamma)^{23}Na}$ resonances  was carried out. Here, new data on the branching ratios of the $E_p$ = 189.5 keV resonance are shown, where discrepancies have been reported in the recent two direct detections \cite{Cavanna15-PRL,Depalo16-PRC,Kelly17-PRC}. Further results on the $\omega \gamma$ and the study of the other  resonances will be reported in a forthcoming publication.

After determining the beam energy of maximum yield by a resonance scan, a high-statistics run was performed at the maximum of the yield curve. The data taken here differ from the previous LUNA data with two HPGe detectors reported in \cite{Cavanna15-PRL,Depalo16-PRC} in three respects: First, the $\gamma$-ray energy resolution is inferior in the new data. Second, the counting rate is hundredfold higher. Third, angle-averaged data are available due to the quasi 4$\pi$ angular coverage of the BGO borehole detector.

\begin{figure}[tb]
\includegraphics[width=\columnwidth]{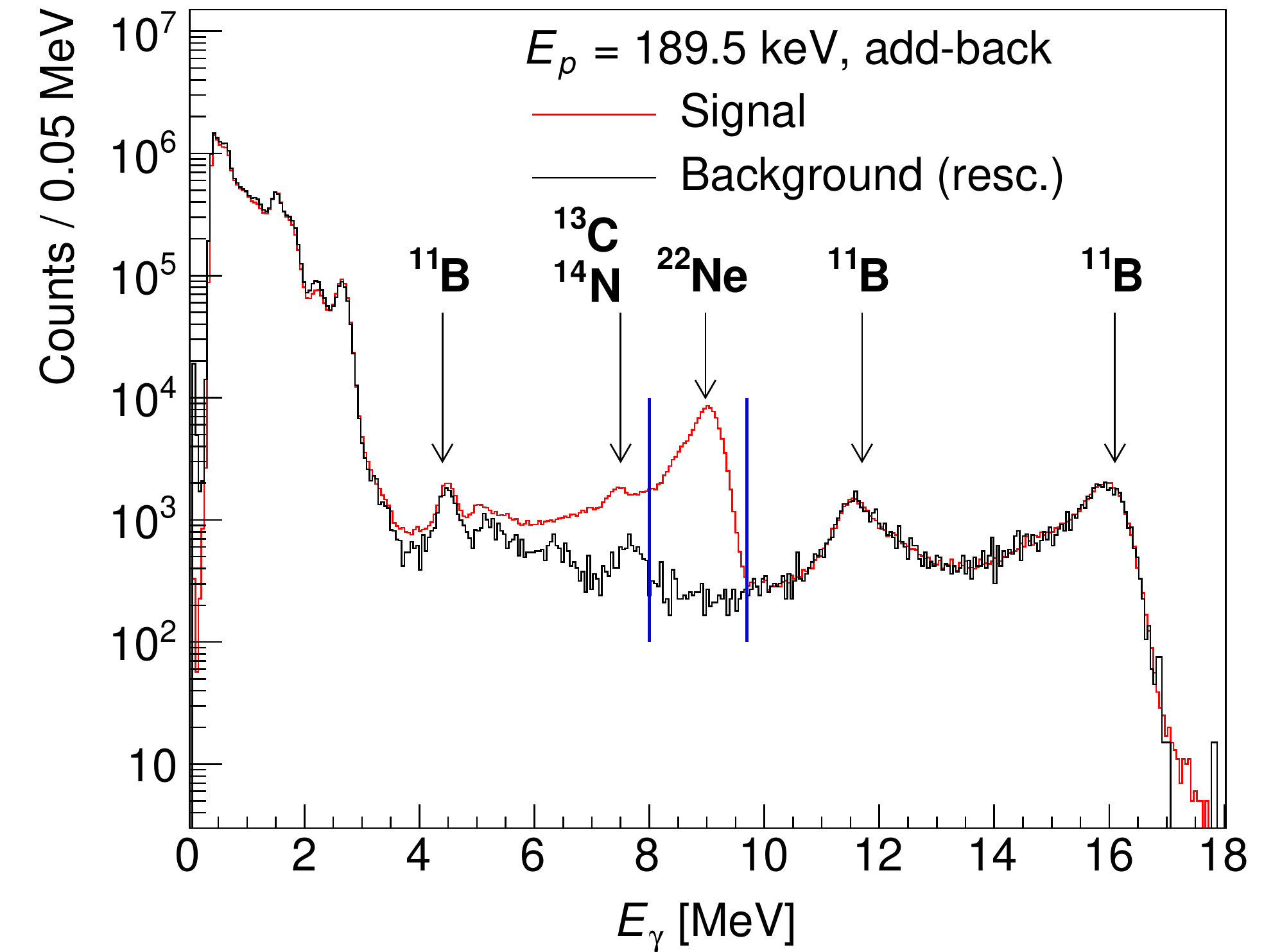}
\caption{\label{fig:figure_new} The experimental spectrum acquired using Neon (Argon) gas in red (black). The two spectra are normalised to match the region $E_\gamma$ = 10.5-17.0\,MeV as discussed in the text. The main sources of beam induced background are also labelled in the figure.}
\end{figure}

In addition to the on-resonance run with a proton beam incident on $^{22}$Ne gas, a background monitor run was performed with proton beam on argon gas. The argon spectrum was used to subtract the beam induced background (given in this instance exclusively by the $\mathrm{^{11}B(p,\gamma)^{12}C}$ reaction), by using the counting rate in the $E_\gamma$ = 10.5-17.0\,MeV region to match the spectra \cite{Takacs17-PhD,Ferraro17-PhD} (see Figure \ref{fig:figure_new}). The background amounted to 7\% of the raw counts in the region of interest (ROI). 

\begin{figure*}[tb]
\includegraphics[width=\textwidth]{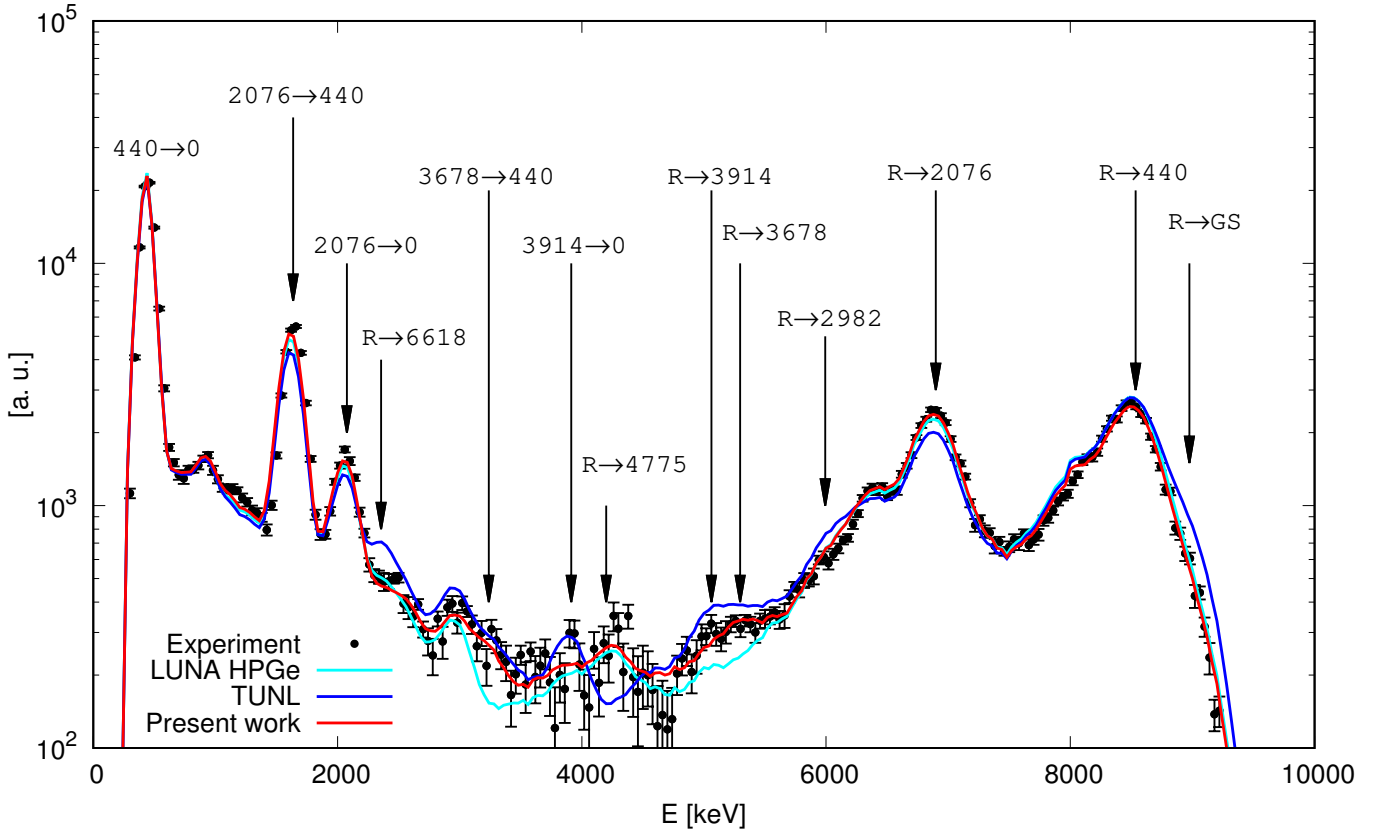}
\caption{\label{fig:Gated189} Single sum spectrum on top of the 189.5\,keV resonance in $^{22}$Ne(p,$\gamma$)$^{23}$Na, gated on the add-back energy in the sum peak, $E_\gamma^{\rm sum}$ $\in$ [8.0;9.7] MeV. The data are compared with simulated branchings from LUNA-HPGe \cite{Depalo16-PRC}, TUNL \cite{Kelly17-PRC}, and the best fit results from the present data, LUNA-BGO.}
\end{figure*}

The $^{22}$Ne(p,$\gamma$)$^{23}$Na ROI in the add-back spectrum was  $E_\gamma$ = 8.0-9.7 MeV. 
As a next step, by gating on the $^{22}$Ne(p,$\gamma$)$^{23}$Na ROI in the add-back spectrum and again subtracting the $^{11}$B background, a $^{22}$Ne(p,$\gamma$)$^{23}$Na only single sum spectrum was generated (Figure \ref{fig:Gated189}). The signatures from the complex decay pattern of the resonance are clearly apparent. 
The appropriateness of the background subtraction is confirmed by the fact that the remaining counts in the ROI can be explained by the $^{22}$Ne(p,$\gamma$)$^{23}$Na resonance. 

\begin{table}[tb]
\begin{tabular}{l c c c c c}
\hline
\noalign{\vskip 1 mm} 
\multicolumn{1}{c}{$\gamma$ transition} & \multicolumn{3}{c}{Branching [\%]}  \\
 & LUNA & TUNL  & LUNA & \\
 & HPGe \cite{Depalo16-PRC} &  \cite{Kelly17-PRC} & BGO [this work] & \\
\noalign{\vskip 1 mm} 
\hline
\hline
\noalign{\vskip 1 mm} 
	8975 $\rightarrow 0$		& 					&	$5.3\pm1.4$			&		$\leq 1$			\\
	8975 $\rightarrow 440$		& $42.8 \pm 0.9$	&	$37.7\pm1.5$		&	$35 \pm 6$	\\
	8975 $\rightarrow 2076$		& $47.9 \pm 0.9$	&	$39.8\pm1.3$			&	$53 \pm 6$	\\
	8975 $\rightarrow 2982$		& $3.7 \pm 0.5$		&	$5.0\pm0.8$		&		$3.3 \pm 0.7$	\\
	8975 $\rightarrow 3678$		& 					&	$2.2\pm0.8$				&	$2.4 \pm 0.5$	\\
	8975 $\rightarrow 3914$		& $1.1 \pm 0.3$		&	$3.1\pm0.6$			&	$1.6 \pm 0.5$	\\
	8975 $\rightarrow 4775$		& $1.8 \pm 0.2$		&	$\leq 3.0$			&	$1.9 \pm 0.4$	\\
	8975 $\rightarrow 6618$		& $2.7 \pm 0.2$		&	$4.7\pm0.9$			&	$2.5 \pm 0.8$	\\
	
	\hline
\hline
\end{tabular}
\caption{\label{tab:Branchings189} Decay branching ratios for the 189.5\,keV resonance in $^{22}$Ne(p,$\gamma$)$^{23}$Na (corresponding to the $E_x$ = 8975\,keV excited state) from LUNA-HPGe \cite{Depalo16-PRC}, TUNL \cite{Kelly17-PRC}, and from the present work, here labeled LUNA-BGO. The work by Jenkins {\it et al.} \cite{Jenkins13-PRC} showed only the decay to the $E_x$ = 2982\,keV level.}
\end{table}

This single sum spectrum was  used to test several hypotheses regarding the branching ratios,  from the LUNA-HPGe experiment \cite{Depalo16-PRC} and from the TUNL experiment \cite{Kelly17-PRC}, both obtained at 55$^\circ$ angle. 
Both sets of branching ratios give a good match at $E_\gamma >$ 6\,MeV and  $E_\gamma <$ 2\,MeV. However, it seems that neither of the two branching sets provides a good match in the central part of the spectrum $E_\gamma$ = 2-6\,MeV.
In addition, the high  branching to the ground state reported in \cite{Kelly17-PRC} seems inconsistent with the experimental spectrum.
The experimental yield at $E_\gamma \sim$ 9\,MeV can be completely explained by summing effects from cascade transitions, and only an upper limit is found for the ground state branch.
 
Despite of the fact that the LUNA BGO summing crystal is not particular sensitive to the branching or gamma cascades, an attempt to determine the branching ratios has been made by fitting the single sum spectrum with the simulated spectra. In Table \ref{tab:Branchings189}, the results of this fitting procedure are shown. The uncertainty on the branching probabilities are based on the errors from the fitting procedure using MINUIT and from cross-checks using simulated templates from both Geant4 and Geant3. It is clear that while any angular effects can be safely excluded, the limited energy resolution limits the precision in the branching values, which remains in general worst than the one reported in \cite{Depalo16-PRC} and \cite{Kelly17-PRC}.

The new LUNA result is in good agreement with the previous one reported by \cite{Depalo16-PRC}. However, thanks to the pretty high efficiency in the new setup, a contribution to the transition to the 3678\,keV level of $^{23}$Na is required. This transition was observed in the HPGe phase with a non-significant number of counts and reported as upper limits ($<$ 0.7\%) in two Ph.D. theses \cite{Depalo15-PhD,Cavanna15-PhD} and then disregarded in the final publication.
This could be an effect of the different coverage of the angular distribution.

When comparing to TUNL, a stronger branching for the main transition, 8975\,keV$\rightarrow$2076\,keV (and onward mainly through the 440\,keV state to the ground state) is found, mainly caused by the observed yield near the $E_\gamma$ = 5899\,keV primary $\gamma$ ray. In contrast, three minor transitions (ground state, 8975\,keV$\rightarrow$2982\,keV, 8975\,keV$\rightarrow$3914\,keV) are found to be weaker.

\section{Summary and outlook}
\label{sec:Summary}

In summary, a new high-efficiency setup was developed to investigate the low energy yield in the $\mathrm{^{22}Ne(p,\gamma)^{23}Na}$, $\mathrm{^{22}Ne(\alpha,\gamma)^{26}Mg}$, and $\mathrm{^{2}H(p,\gamma)^{3}He}$ reactions. The setup has been characterised and tested.

As a first application of the new setup, the decay branching ratios of the $E_p$ = 189.5 keV resonance in \linebreak $^{22}$Ne(p,$\gamma$)$^{23}$Na were determined, independently from any possible angular effect.

The new results confirm the previous LUNA ones \cite{Depalo16-PRC} and show a stronger branching at 8975\,keV$\rightarrow$2076\,keV with respect to  TUNL \cite{Kelly17-PRC} while do not confirm the evidence of a transition to the ground state in the $E_p$ = 189.5 keV resonance. 

This result also shows the capability of the new setup for determining branching ratios even with complex cascade transition and using a nearly 4$\pi$ detection setup with moderate energy resolution. This could be important when investigating resonances with very low resonance strength, not detectable with detection system of lower  efficiency.

\subsection*{Acknowledgments}

Financial support by INFN, by the Helmholtz Association Nuclear Astrophysics Virtual Institute (NAVI, HGF VH-VI-417), by the NKFIH K120666, by the Hungarian Academy of Sciences (LP2014-17), and by the Deutsche Forschungsgemeinschaft (DFG, BE 4100/4-1)  is gratefully acknowledged.



\begin{thebibliography}{34}

\bibitem{Iliadis15-Book}
C.~Iliadis, \emph{Nuclear Physics of Stars}, {2$^{\rm nd}$}~edn. (Wiley-VCH,
  Weinheim, 2015)

\bibitem{Gratton04-ARAA}
R.~{Gratton}, C.~{Sneden}, E.~{Carretta}, Annual Review of Astronomy and Astrophysics \textbf{42}, 385 (2004)

\bibitem{Longland10-PRC}
R.~{Longland}, C.~{Iliadis}, J.M. {Cesaratto}, A.E. {Champagne}, S.~{Daigle},
  J.R. {Newton}, R.~{Fitzgerald}, Phys.~Rev.~C \textbf{81}, 055804 (2010)

\bibitem{Depalo15-PRC}
R.~Depalo, F.~Cavanna, F.~Ferraro, A.~Slemer, T.~Al-Abdullah, S.~Akhmadaliev,
  M.~Anders, D.~Bemmerer, Z.~Elekes, G.~Mattei et~al., Phys. Rev. C
  \textbf{92}, 045807 (2015)

\bibitem{NACRE99-NPA}
C.~{Angulo}, M.~{Arnould}, M.~{Rayet}, P.~{Descouvemont}, D.~{Baye},
  C.~{Leclercq-Willain}, A.~{Coc}, S.~{Barhoumi}, P.~{Aguer}, C.~{Rolfs}
  et~al., Nucl.~Phys.~A \textbf{656}, 3 (1999)

\bibitem{Iliadis10-NPA841_31}
C.~{Iliadis}, R.~{Longland}, A.E. {Champagne}, A.~{Coc}, R.~{Fitzgerald},
  Nucl.~Phys.~A \textbf{841}, 31 (2010)

\bibitem{Cavanna15-PRL}
F.~{Cavanna}, R.~{Depalo}, M.~{Aliotta}, M.~{Anders}, D.~{Bemmerer}, A.~{Best},
  A.~{Boeltzig}, C.~{Broggini}, C.G. {Bruno}, A.~{Caciolli} et~al.,
  Phys.~Rev.~Lett. \textbf{115}, 252501 (2015)

\bibitem{Depalo16-PRC}
R.~Depalo, F.~Cavanna, M.~Aliotta, M.~Anders, D.~Bemmerer, A.~Best,
  A.~Boeltzig, C.~Broggini, C.G. Bruno, A.~Caciolli et~al. (LUNA
  Collaboration), Phys. Rev. C \textbf{94}, 055804 (2016)

\bibitem{Slemer17-MNRAS}
A.~Slemer, P.~Marigo, D.~Piatti, M.~Aliotta, D.~Bemmerer, A.~Best, A.~Boeltzig,
  A.~Bressan, C.~Broggini, C.G. Bruno et~al., Monthly Notices of the Royal
  Astronomical Society \textbf{465}, 4817 (2017)

\bibitem{Kelly17-PRC}
K.J. Kelly, A.E. Champagne, L.N. Downen, J.R. Dermigny, S.~Hunt, C.~Iliadis,
  A.L. Cooper, Phys. Rev. C \textbf{95}, 015806 (2017)

\bibitem{Cavanna14-EPJA}
F.~{Cavanna}, R.~{Depalo}, M.L. {Menzel}, M.~{Aliotta}, M.~{Anders},
  D.~{Bemmerer}, C.~{Broggini}, C.G. {Bruno}, A.~{Caciolli}, P.~{Corvisiero}
  et~al., Eur.~Phys.~J.~A \textbf{50}, 179 (2014)

\bibitem{Formicola03-NIMA}
A.~{Formicola}, G.~{Imbriani}, M.~{Junker}, D.~{Bemmerer}, R.~{Bonetti},
  C.~{Broggini}, C.~{Casella}, P.~{Corvisiero}, H.~{Costantini}, G.~{Gervino}
  et~al., Nucl.~Inst.~Meth.~A \textbf{507}, 609 (2003)

\bibitem{Costantini09-RPP}
H.~Costantini, A.~Formicola, G.~Imbriani, M.~Junker, C.~Rolfs, F.~Strieder,
  Reports on Progress in Physics \textbf{72}, 086301 (2009)

\bibitem{Broggini10-ARNPS}
C.~Broggini, D.~Bemmerer, A.~Guglielmetti, R.~Menegazzo, Annu. Rev. Nucl. Part.
  Sci. \textbf{60}, 53 (2010)

\bibitem{Broggini18-PPNP}
C.~Broggini, D.~Bemmerer, A.~Caciolli, D.~Trezzi, Prog. Part. Nucl. Phys. \textbf{98}, 55 (2018)

\bibitem{Ziegler10-NIMB}
J.F. Ziegler, M.D. Ziegler, J.P. Biersack, Nucl.~Inst.~Meth.~B \textbf{268},
  1818 (2010)

\bibitem{Caciolli12-EPJA}
A.~{Caciolli}, D.A. {Scott}, A.~{Di Leva}, A.~{Formicola}, M.~{Aliotta},
  M.~{Anders}, A.~{Bellini}, D.~{Bemmerer}, C.~{Broggini}, M.~{Campeggio}
  et~al., Eur.~Phys.~J.~A \textbf{48}, 144 (2012)

\bibitem{Bordeanu13-NPA}
C.~{Bordeanu}, G.~{Gy{\"u}rky}, Z.~{Hal{\'a}sz}, T.~{Sz{\"u}cs}, G.G. {Kiss},
  Z.~{Elekes}, J.~{Farkas}, Z.~{F{\"u}l{\"o}p}, E.~{Somorjai}, Nucl.~Phys.~A
  \textbf{908}, 1 (2013)

\bibitem{Goerres80-NIM}
J.~{G\"orres}, K.~{Kettner}, H.~{Kr\"awinkel}, C.~{Rolfs}, Nucl.~Inst.~Meth.
  \textbf{177}, 295 (1980)

\bibitem{Marta06-NIMA}
M.~{Marta}, F.~{Confortola}, D.~{Bemmerer}, C.~{Boiano}, R.~{Bonetti},
  C.~{Broggini}, M.~{Casanova}, P.~{Corvisiero}, H.~{Costantini}, Z.~{Elekes}
  et~al., Nucl.~Inst.~Meth.~A \textbf{569}, 727 (2006)

\bibitem{Osborne84-NPA}
J.~Osborne, C.~Barnes, R.~Kavanagh, R.~Kremer, G.~Mathews, J.~Zyskind,
  P.~Parker, A.~Howard, Nuclear Physics A \textbf{419}, 115  (1984)

\bibitem{Vlieks83-NIM}
A.~Vlieks, M.~Hilgemeier, C.~Rolfs, Nuclear Instruments and Methods in Physics
  Research \textbf{213}, 291  (1983)

\bibitem{Casella02-NIMA}
C.~{Casella}, H.~{Costantini}, A.~{Lemut}, B.~{Limata}, D.~{Bemmerer},
  R.~{Bonetti}, C.~{Broggini}, L.~{Campajola}, P.~{Cocconi}, P.~{Corvisiero}
  et~al., Nucl.~Inst.~Meth.~A \textbf{489}, 160 (2002)

\bibitem{Ferraro17-PhD}
F.~Ferraro, Ph.D. thesis, Universit\`a degli Studi di Genova (2017)

\bibitem{Caciolli11-AA}
A.~{Caciolli}, C.~{Mazzocchi}, V.~{Capogrosso}, D.~{Bemmerer}, C.~{Broggini},
  P.~{Corvisiero}, H.~{Costantini}, Z.~{Elekes}, A.~{Formicola},
  Z.~{F{\"u}l{\"o}p} et~al., Astron.~Astrophys. \textbf{533}, A66 (2011)

\bibitem{Boeltzig17-JOP}
A.~Boeltzig et~al., J. Phys. G \textbf{45}, 025203 (2018)

\bibitem{ROOT}
R.~Brun, F.~Rademakers, Nuclear Instruments and Methods in Physics Research
  Section A: Accelerators, Spectrometers, Detectors and Associated Equipment
  \textbf{389}, 81  (1997), new Computing Techniques in Physics Research V

\bibitem{Bemmerer05-EPJA}
D.~Bemmerer, F.~{Confortola}, A.~{Lemut}, R.~{Bonetti}, C.~{Broggini},
  P.~{Corvisiero}, H.~{Costantini}, J.~{Cruz}, A.~{Formicola},
  Z.~{F{\"u}l{\"o}p} et~al., Eur.~Phys.~J.~A \textbf{24}, 313 (2005)

\bibitem{Lemut06-PLB}
A.~{Lemut}, D.~{Bemmerer}, F.~{Confortola}, R.~{Bonetti}, C.~{Broggini},
  P.~{Corvisiero}, H.~{Costantini}, J.~{Cruz}, A.~{Formicola},
  Z.~{F{\"u}l{\"o}p} et~al., Phys.~Lett.~B \textbf{634}, 483 (2006)
  
\bibitem{Bemmerer06-NPA}
D.~Bemmerer, F.~{Confortola}, A.~{Lemut}, R.~{Bonetti}, C.~{Broggini},
  P.~{Corvisiero}, H.~{Costantini}, J.~{Cruz}, A.~{Formicola},
  Z.~{F{\"u}l{\"o}p} et~al., Nucl.~Phys.~A \textbf{779}, 297 (2006)

\bibitem{Bemmerer09-JPG}
D.~{Bemmerer}, A.~{Caciolli}, R.~{Bonetti}, C.~{Broggini}, F.~{Confortola},
  P.~{Corvisiero}, H.~{Costantini}, Z.~{Elekes}, A.~{Formicola},
  Z.~{F{\"u}l{\"o}p} et~al., J.~Phys.~G \textbf{36}, 045202 (2009)

\bibitem{Takacs17-PhD}
M.~Tak\'acs, Ph.D. thesis, Technische Universit\"at Dresden and Helmholtz
  Zentrum Dresden Rossendorf (2017)

\bibitem{Jenkins13-PRC}
D.~Jenkins, M.~Bouhelal, S.~Courtin, M.~Freer, B.~Fulton et~al., Phys.Rev. C
  \textbf{87}, 064301 (2013)

\bibitem{Depalo15-PhD}
R.~Depalo, Ph.D. thesis, {Universit\`a degli Studi di Padova} (2015)

\bibitem{Cavanna15-PhD}
F.~Cavanna, Ph.D. thesis, {Universit\`a degli Studi di Genova} (2015)

\end{thebibliography}


\end{document}